\documentclass{latex-template}

\usepackage{amsmath}
\usepackage{amssymb}
\usepackage{abstract} 
\usepackage[english]{babel} 
\usepackage{booktabs} 
\usepackage[hang, small,labelfont=bf,up,textfont=it,up]{caption} 
\usepackage{enumitem} 
\usepackage{graphicx}
\usepackage{hyperref} 
\usepackage[utf8]{inputenc}
\usepackage{listings, lstautogobble} 
\usepackage{url} 

\begin{document}
\title{
Gophy: Novel Proof-of-Useful-Work blockchain architecture for High Energy Physics
\thanks{
Work supported by HFHF
}}

\author{Felix Hoffmann\thanks{felix.hoffmann@iri.uni-frankfurt.de}}
\affil{FIAS, Frankfurt am Main, Germany}
\author{Udo Kebschull\thanks{uk@rz.uni-frankfurt.de}}
\affil{Goethe University Frankfurt, Germany}
\maketitle

\begin{abstract}
	In this publication, a novel architecture for Proof-of-Useful-Work blockchain consensus which aims to replace hash-based block problems with Monte Carlo simulation-based block problems to donate computational power to real-world HEP experiments is described. 
	Design decisions are detailed and challenges are addressed. The architecture is being implemented using Golang and can be run inside the CbmRoot software environment. The goal is to build a bridge between the disciplines HEP and blockchain to build a novel blockchain network in which the network's computational power is not wasted but instead used to support a scientific experiment while at the same time securing the underlying permissioned blockchain. The blockchain features a token-based cryptocurrency that is rewarded to miners that donate computational power and acts as an additional incentive to participate which traditional volunteer computing can not provide. The implementation named gophy is being implemented in Golang and is expected to be open-sourced before the end of 2024.
\end{abstract}

\section{Introduction}
Hash-based Proof-of-Work (PoW) requires difficult block problems and is commonly criticized for using large amounts of energy and computational power without creating benefit outside the blockchain. At the same time, High Energy Physics experiments all around the world rely on computationally expensive Monte Carlo simulations for tasks such as R\&D of new detectors, optimization of existing ones, background subtraction and interpretation of real data. The intention of the authors is to bring blockchain and HEP communities together: We are working on a novel architecture for enabling Proof-of-Useful-Work (PoUW) blockchain consensus that will support the scientific HEP experiment CBM (Compressed Baryonic Matter) at GSI/FAIR with MC simulation data while securing the underlying blockchain. More specifically, instead of solving hash-based block problems miners will run requested Monte Carlo simulations in CbmRoot \cite{cbmroot}, which is the software environment used by the experiment. In order to preserve the usefulness aspect, a permissioned node controlled by experiment insiders is responsible for defining new block problems and collecting results. In the following, this node will be referred to as the \textit{Root Authority} (RA). In regular time intervals, the RA will publish well-defined problem definitions that instruct miners with access to CbmRoot to run MC simulations of events that currently are of interest to the experiment. Each miner receives the exact same problem definition and everyone uses the same RNG seed (which is part of the problem definition) so that results are reproducible and comparable. A problem is solved by running the specified simulation and uploading resulting data to the RA. Gophy implements communication between nodes by making use of the libp2p network.\\\\
The advantage of the proposed novel blockchain network over traditional volunteer computing approaches (such as e.g. the BOINC project LHC@home \cite{lhcAtHome}) is that an incentive to participate and donate computational power can be provided in the form of tradable tokens that are generated and rewarded to the winner node with each new block.\\
While not the focus of this publication, more details about the theory behind Proof-of-Useful-Work and the challenges of transitioning from hash-based PoW to PoUW can be found in linked literature. \cite{hoffmannChallenges}

\section{Architecture}
The blockchain network consists of peers that communicate with each other (e.g. requesting block information for syncing) and a RA which is responsible for defining new block problems, collecting results and determining the block winner. In order to limit the power of the RA, several precautions have been taken when designing the blockchain architecture so that it is not built on trust, but instead trust is created by equipping every node with the necessary verification tools. This means that every participant is able to control the actions of the RA and determine whether it acts according to the rules everyone has agreed on. For instance, the block winner selection algorithm is designed to be fair, transparent, reproducible and unpredictable which makes it impossible for the RA to favor one miner over another while allowing every blockchain participant to verify the winner selection outcome. Details about the algorithm can be found in \cite{dftws}.\\\\
For node discovery and communication between peers libp2p \cite{libp2p} with protocols like Kademlia-DHT, Rendezvous and PubSub (Publish-Subscribe) is being used. Direct communication between nodes is also possible via a chat protocol which is used for uploading solution data directly to the RA. Every message is signed using Ed25519, which means that it is not possible to impersonate the RA (whose public key and node ID is known) to send out fake problems to the network.

\subsection{Database structure}
In the official client software implementation named \textit{gophy}, which is written in Go, a NoSQL key-value database with bucket support called bbolt \cite{bbolt} is used to store block data as key-value pairs. A common pattern is to store data as BlockHash-SerializedBlockContent pairs, which allows for efficient retrieval of specific blocks. For serialization and deserializing gophy uses msgpack \cite{msgpack}, which is a Go wrapper for the binary serialization format MessagePack. In the context of blockchain software, NoSQL databases are commonly preferred over SQL alternatives because a block can be identified by its hash and by having each block reference the previous block hash a natural blockchain traversal mechanism is automatically created.\\\\
Within the bbolt database, separate buckets are used to hold different kinds of data:
\begin{itemize} 
	\item Chain db bucket\hyperref[tab:chaindb_structure]{[Table 1]} (Contains block data such as e.g. list of transactions or problem definition parameters)
	\item State db bucket\hyperref[tab:statedb_structure]{[Table 3]} (Used to keep track of account token balances)
\end{itemize}
In the following more information about the purpose of each bucket is given:
\subsubsection{Chain db bucket}
The chain db bucket is synced between all nodes and contains the data of blockchain blocks. A block consists of a header and additional data such as list of transactions and a problem definition. Gophy supports multiple sync modes which means that full nodes store entire blocks, but there also exist light nodes that only store headers.\\\\
Below shows the simplified structure of chain db:\\
\begin{center}
\begin{table}[!htbp]	
	\centering 
	\caption{Chain db structure}
	\label{tab:chaindb_structure}
	\begin{tabular}{ccl}
		\toprule
		Key	& Value\\
		\midrule
		``\verb|latest|''		& Block Key \\
		Block Key				& Serialized Block Content \\
		Block Key				& Serialized Block Content \\
		...				& ... \\
		\bottomrule
	\end{tabular}
\end{table}
\end{center}
Each row with key ``\verb|Block Key|'' represents a block of the blockchain. In order to enable a simple mechanism to access specific blockchain data, the key of a block is equal to the Keccak256 hash of its header. When syncing data with other nodes, any attempt at block manipulation (changed bit in the value field) would lead to a different header which leads to a different key for the block, which can be detected.\\
To efficiently access the most recent block, there exists a special row with key ``\verb|latest|''. Its key is a constant string and its value is the key of the newest block. As a result, it must be updated whenever a new block is added to the chaindb bucket. This key-value pair enables blockchain traversal: ``\verb|Latest|'' is accessed to get the newest block and from there on each block references the previous block.\\\\
Here is an overview of which data a chain db block holds:
\begin{center}
\begin{table}[!htbp]	
	\centering
	\caption{Chain db block content}
	\label{tab:blockContent}
	\begin{tabular}{||c||}
		\toprule
		Key: Hash of Header \\
		\midrule
		\textbf{Header:} \\
		1. BlockTime \\
		2. BlockID \\
		3. PrevBlockHash \\
		4. TransactionsMerkleRoot \\
		5. StateMerkleRoot \\
		6. ProblemID \\
		7. BlockWinner \\\\
		\textbf{Full node only:} \\
		8. Transactions \\
		9. BlockProblemDefinition \\ [1ex]
		\hline
		\bottomrule
	\end{tabular}
\end{table}
\end{center}
In the following, details about what the purpose of each entry is are given:\\\\
0. \textit{Block Hash}: The block hash is determined by $Keccak256(serializedHeader)$. Only the header is hashed so that full and light nodes agree on the same block hash. The header contains certain Merkle tree root hash data that makes it possible for light nodes to request data from full nodes while being able to verify that the received data was not manipulated and is in fact part of the block.\\\\
1. \textit{BlockTime}: The timestamp is used to keep track of when a block was created. There are certain block transition rules that determine whether $Block(x+1)$ can be a continuation of $Block(x)$. While details are outside the scope of this paper, it should be noted that the timestamp is used as one of many factors.\\\\
2. \textit{BlockID}: Block IDs are used to make traversing the blockchain more human friendly. After block $x$ block $x+1$ follows, etc. The IDs play a role in the equation that determine the amount of tokens that are to be rewarded for a block, because the awarded amount is halved in regular intervals.\\\\
3. \textit{PrevBlockHash}: The hash of the previous block is an important value that must be included in each block because it allows maintaining the integrity, security and immutability of the blockchain. A blockchain by definition is a chain of blocks. The chain is only formed if, for any block, the previous block is known (except for the block with BlockID 0, which is called the genesis block). If any data in any previous block were tampered with, then all following block hashes would change. This mechanism allows the implementation of a sync protocol where nodes can request blockchain data from other nodes and verify their validity. If two nodes agree on the hash of the previous block, then this implies that they agree on everything that has happened on the blockchain until that block.\\
This value is also used by the RA to determine the RNG seed of the next problem definition. Note that the simulation seed does not affect the usefulness of results. Monte Carlo simulations in HEP model probabilistic processes according to Standard Model expectations. There is no reason to prefer one RNG seed over another one, even though it will affect the result itself, because their statistical distribution is in agreement with Standard Model expectations. In the context of this publication, a \textit{correct} solution does not mean using a different RNG seed would have led to a less desirable outcome. Instead, it means that the RA defined the problem to be running a simulation with a given set of parameters (which includes a specific RNG seed). If a miner used a different RNG seed, then the results will be considered to be invalid in the context of this blockchain because the probabilistic algorithm that determines the accepted solution partly relies on the assumption that the most common solution is more likely to be the correct solution. This algorithm is described in more detail in section \hyperref[sec:verification]{Verification of results}. This is why it is mandatory that every miner uses the same RNG seed for solving a given block problem.\\\\
4. \textit{TransactionsMerkleRoot}: The transaction list itself is only locally stored by full nodes, but light nodes are able to verify the validity of data they request and receive from full nodes. Assume a scenario in which a light node $l$ is interested in details of a transaction with unique ID $t$. First, $l$ requests the data $t$ contains from full nodes, who retrieve it and send it back to $l$, along with a Merkle proof. In order to allow $l$ to verify that the received data was not manipulated, all transactions IDs (which are calculated by hashing the transactions contents) are used as leaves in a Merkle tree which is constructed bottom-up (concatenate two hashes, then hash them to create parent hash). As a result, there will be one hash at the top of the tree (root hash) which is stored as TransactionsMerkleRoot in the header (that the light node has access to because it is stored in the header of a block). $l$ can now use the Merkle proof (which contains sibling node hashes) to verify that the received piece of data was part of the tree that led to the known root hash. This mechanism allows for a lightweight sync protocol in which light node only need to know the root hash of the Merkle tree in order to verify the validity of any data whose hashes it contains.\\\\
5. \textit{StateMerkleRoot}: The state db bucket keeps track of the current token balance and transaction nonce of each node. Its entries form the leaves of a Merkle Tree which then is constructed bottom-up to determine the root hash of the state db at this point in time. By including the root hash of the state db bucket in the header of a chain db block, a light node mechanism to request verifiable information about any node's token amount or transaction nonce becomes possible.\\\\
6. \textit{ProblemID}: Each block has an underlying block problem defined by the RA that must be solved by miners. Details of the block problem are only stored by full nodes, but its hash is part of the block header are therefore also known by light nodes. This not only makes it possible to refer to a specific block problem with a single hash but also binds the problem to a specific block.\\\\
7. \textit{BlockWinner}: From all miners that solved the block problem, the RA chooses one block winner and awards a certain amount of tokens to it. The algorithm used for this purpose is described in detail in linked literature \cite{dftws}. It is important to include the address of the winner node in the block itself. This is required so that any full node can locally build the wallet state of each node at any point in time.\\
The advantage of a blockchain-based approach compared to traditional volunteer computing is that it can provide an incentive to participate by rewarding tradable tokens. Tokens can be seen as the underlying currency of the blockchain that depending on public perception might or might not have value associated with them. There currently are no plans to implement a mechanism which allows nodes to exchange their tokens for real-world currency, but it technically could be implemented, if the RA would be willing to provide such a service.\\\\
8. \textit{Transactions}: Each block contains a list of transactions. Any node is able to broadcast transactions which allows the transfer of tokens (that are rewarded for solving block problems) between nodes. Each broadcast transaction must be signed using Ed25519 to prevent impersonation of nodes. A transaction nonce is used (and incremented by one for each transaction a node has successfully performed) to prevent replay attacks: It is not possible for a node to re-broadcast an old signed transaction of a different node to re-trigger it because either the nonce would not be changed which can be detected (because the state can locally be created at any point in time), or the nonce is changed which invalidates the signature which also can be detected. The RA decides which transactions are included in a block, but there are rules in place, that are outside of the scope of this paper, to prevent the RA from disallowing or ignoring unwanted transactions.\\\\
9. \textit{BlockProblemDefinition}: Each block problem consists of a header (which contains metadata such as previous block hash, problem expiry timestamp, etc.) and a list of simulation parameters that specifies which simulation the miners need to run to perform useful work and be eligible for winner selection. Each problem is bound to a block by referencing the previous block hash. The MC simulation problem can be represented by a simple list of parameters that affect event generation, Geant \cite{geant} configurations or other aspects of the simulation setup. The following list contains a few parameters that could be part of a block problem definition:
\begin{itemize}\label{tab:blockproblem_parameters}
	\item RNG seed (the only parameter that the RA can not arbitrarily choose, it has to be derived from PrevBlockHash)
	\item Luminosity
	\item Amount of events
	\item Run ID
	\item Particle types
	\item Momentum
	\item Theta
	\item Choice of Geant physics list
	\item QCD scale
	\item Background processes
	\item Active detector list
\end{itemize}
It is important that any code involved with running the simulation is designed in a way that allows the setting of custom RNG seeds to allow reproducibility of results. After receiving the simulation parameters, miners pass them to a C script in their CbmRoot environment which coordinates the setup and execution of the simulation task. The RA ideally is able to pass new versions of this script via the block problem definition so that they are flexible to modify future block problems like they desire.
\subsubsection{State db bucket}
The state db bucket is not synced between nodes but instead locally built by each full node with the data from the chaindb bucket. The state db bucket is used to store nodeID-nodeWallet pairs so that nodes can keep track of all token balances and transaction nonces of other nodes, which is necessary to efficiently determine the validity of a given transaction: Each chain db block contains a list of transactions, but in order to determine whether the block is valid or not a node must be able to verify that the Sender node's balance was equal-or-larger than the sum of the token balance it wants to send to a Receiver node and the transaction fee it is willing to burn (used by RA to decide which transactions are to be prioritized for block inclusion as there is a maximum amount of transactions that can be included per block). In addition to the token balance, each state db value also keeps track of the node's transaction nonce, which acts as a counter of sent transactions that is increased by one with each transaction. As previously mentioned, this is necessary to prevent replay attacks in which a malicious node re-broadcasts signed transactions that had already been included in a previous block to spend the same amount of tokens again. By forcing a nonce to be included in every transaction, nodes will be able to detect such malicious attempts.\\
If a node has access to the chain db bucket, then the state db bucket can locally be re-created at any point in time by chronologically traversing the blockchain and keeping track of which transactions and block rewards occurred. Light nodes are able to request e.g. current token balances of other nodes via topic requests and can verify the received information and Merkle proof by referring to the StateMerkleTreeRootHashes saved in locally stored chaindb block headers. This allows light nodes to request and verify certain data without being forced to locally store all blockchain data, which lowers the entry barrier to join the blockchain network. It should be noted that while miners must have access to CbmRoot, other nodes can run gophy on any OS that supports Golang without needing access to software contained in CbmRoot.\\\\
The state db contains a special row with key ``\verb|latest|'' that stores which point in time the current state of the bucket describes: The value of this row is the hash of the latest chaindb block at that point in time, which can be used to retrieve the data of the latest block from the chaindb, which contains the exact timestamp along with other block information.\\\\
The following shows a simplified structural view of the state db:
\begin{center}
\begin{table}[!htbp]
	\centering
	\caption{State db structure}
	\label{tab:statedb_structure}
	\begin{tabular}{ccl}
		\toprule
		Key					& Value\\
		\midrule
		``\verb|latest|'' 	& Latest chain db key \\
		Node address (hex)	& TokenAmount, Nonce   \\
		Node address (hex)	& TokenAmount, Nonce   \\
		...					& ... \\
		\bottomrule
	\end{tabular}
\end{table}
\end{center}
\subsection{Message distribution}
As previously mentioned, libp2p is used for node discovery and exchange of data between peers. It features PubSub implementations that allow for efficient message distribution via topics. The basic idea is that a node can subscribe to a \textit{topic} which makes it receive all messages sent to that topic. Vice versa, any node can send messages to a topic in order to reach all nodes that have subscribed to this topic.\\\\
Gophy currently makes use of the following topics:
\begin{enumerate}
	\item Chaindb topic
	\item Transactions topic
	\item Block problem topic
	\item Block creation topic
	\item Miner commitment topic
\end{enumerate}
In the following, the purpose of each of these topics is briefly described:\\\\
1. \textit{Chaindb topic}: This topic is used by nodes to sync blockchain data (anything contained in the chaindb bucket) without having to rely on the RA to be reachable. If any new node joins the blockchain network, it can send requests for any block or a list of all block hashes to this topic. The actual responses to such requests happen via direct communication (libp2p chat protocol) to avoid flooding the topic.\\\\
2. \textit{Transactions topic}: In this topic, transaction intentions are broadcast. A transaction itself holds information about how many tokens are transferred from a sender node to a receiver node, the correct sender transaction nonce, the fee the sender is willing to pay for prioritized block inclusion, etc. This information is hashed to get a unique transaction ID, and that hash is signed by the sender node. Finally, the data can be serialized and sent to the transaction topic where it is pending until the RA creates a new block that includes it (expiration dates of transaction could be added to invalidate a transaction if it has not been included in a block until a certain point in time). The RA prioritizes pending transactions with the highest fees (fees are burned, the RA does not collect them) and every node is able to control that the RA follows this rule because transaction intentions are publicly broadcast to the topic.\\\\
3. \textit{Block problem topic}: This topic is only used by the RA (as in the RA is supposed to be the only sender of messages) to broadcast new block problem tasks. Miners listen to this topic and start running the received simulation tasks, but only if it has been signed by the RA. If they are able to solve the problem and reply to the RA with the correct solution before the problem expires, they are eligible for being selected block winner. Miners will ignore any data in this topic that is not signed by the RA. Each problem definition contains the current timestamp (which can not be lower than the timestamp of any existing chaindb block) and expiry date which defines until when the problem must be solved. This makes it infeasible for malicious nodes (as in technically possible, but every node will be able to detect it) to re-broadcast old, correctly signed problem definitions to try to fool other miners. The RA will target a block time of around 24 hours to be able to send out complex simulation problems while also being able to provide a reasonably steady transaction throughput.\\\\
4. \textit{Block creation topic}: This topic is only used by the RA to broadcast newly created chaindb blocks. As soon as the current block problem has expired and a block winner has been chosen, the RA creates and signs a new chaindb block. This block then is sent via this topic. Nodes ignore any incoming data that is not signed by the RA. Due to the transparent design of the blockchain structure, every node is able to verify that the block sent out by the RA has been created according to the rules every participant agrees on.\\\\
5. \textit{Miner commitment topic}: The algorithm described in \cite{dftws} relies on cryptographic commitments from both the RA and miners to select a block winner. After solving the block problem, miners will broadcast a commitment to this topic which (retrospectively) will prove that they knew the solution at this point in time without revealing it. The idea is based on zero knowledge proofs, and it is used to allow for a transparent block winner selection process without revealing to other nodes any information that would help them to solve the block problem.

\subsection{Reproducible block problems}
As shown in the list of \hyperref[tab:blockproblem_parameters]{problem definition parameters}, a well-defined problem is a list of parameters that need to be passed to a script in CbmRoot to run a set of reproducible Monte Carlo simulations with a custom RNG seed. Forcing the RNG seed to be derived from the previous block hash makes it possible to have reproducible results (each node works on the exact same problem) while also enabling consensus between nodes which seed is to be used. It should be noted that each block problem in fact is a set of many sub-problems (each with their own list of parameters) that can be solved independently. By having reproducible results a probabilistic solution verification algorithm can be used to determine the most common solution (verification-by-replication). Additionally, the RA is able to pre-calculate the results of randomly chosen sub-problems to compare each miner's solution with known correct results. By combining both approaches it is possible to probabilistically filter out incorrect solutions which is a crucial step of the solution verification algorithm described in section \hyperref[sec:verification]{Verification of results}.\\
Should it turn out that floating-point operation differences on different hardware using the same software environment will be problematic in practice, then the code will be adjusted to use a similarity-based clustering approach instead of exact hash matching to determine the most common solution. While the most commonly uploaded solution is used as a factor to determine what likely is the correct solution of the underlying block problem, it is not the only mechanism to detect malicious nodes that try to upload false data, as described in the linked section.\\\\
As previously mentioned, the RNG seed choice does not affect the usefulness of results. Event generators such as e.g. Pythia \cite{pythia} (version 8.3) support setting custom seeds:
\begin{lstlisting}[language=go]
pythia8->ReadString("Random:
setSeed=on");
pythia8->ReadString("Random:
seed=42");
\end{lstlisting}
By using these commands in the simulation script, the event generation process produces the exact same output for repeated runs that use the same parameters. However, event generation is only one aspect of a bigger simulation pipeline. Simulating the passage of particles through detector material, digitization and track reconstruction are tasks that can be part of the block problem as well. It has to be ensured that the software environment is set up in a way that support the setting of a custom RNG seed for every step in this simulation pipeline to guarantee reproducible results. As a result, switching between different software environments is non-trivial because certain parts of the code might have to be adjusted and recompiled to support custom seeds in every step of the simulation pipeline.\\\\
The output of resulting simulation data can be stored as ROOT binary files or as HepMC \cite{hepmc} records. The former has a binary mode that specifically is designed to have consistent hashes by not being affected by metadata such as creation timestamps that are used within ROOT files. When defining the output ROOT file, one must add
\begin{lstlisting}[language=go]
?reproducible=<fixedname>
\end{lstlisting}
to the code to enforce this behavior. \cite{RootReproducible}\\
The latter (output as HepMC records) has the advantage that Rivet \cite{rivet} supports it, which means that part of the problem itself could be running an analysis task on the resulting simulation data. This approach could help to reduce the amount of data that has to be transferred to the RA as the solution itself could in this case be a few resulting plots. Furthermore, the usefulness of results aspect would be strengthened as resulting plots could be added to the MCPlots \cite{mcplots} online repository. MCPlots not only contains comparisons between Standard Model expectations and real-world experimental data but also comparisons between different versions of (possible the same) event generator software. This makes it a viable option for early block problems for the CBM experiment which, as of writing, is still being constructed.

\subsection{Verification of results}\label{sec:verification}
After the block problem expires and the RA has collected the miner solutions, the next step is to determine which solution hash is accepted to be correct solution. Note that the RA does not know the solution of every sub-problem, but it has pre-calculated the solution of at least one randomly chosen sub-problem. The following two approaches are combined to probabilistically try to filter out false data:\\\\
The first filtering stage is \textit{sub-problem matching}. The RA has already solved a random subset of the block problem (it is crucial that nodes can not know or predict which sub-problem solutions are known by the RA) to create points of reference. In this filtering stage, the RA compares each received miner solution with the known sub-problem solutions. This works because the simulation tasks are reproducible. If the RA determines that a miner has uploaded a solution that contains a sub-problem that does not pass this test, the entire solution is deemed false and that node will not be eligible for the winner selection process and the miner is excluded from the next filtering stage.\\\\
The second filtering stage is determining the most-common solution of miners that passed the first filtering stage. Here, instead of viewing each sub-problem separately the RA only compares the hashes of the entire solution (sub-problems are combined and hashed to create one value that represents the entire solution data). The RA loops over each miner solution hash and keeps track of how often a solution with this hash has been uploaded by different miners. The most common solution hash will be regarded as the accepted solution hash. As soon as the accepted solution hash has been determined, the RA puts each miner that uploaded a solution with this hash on the winner selection list. This means that any miner who passes both filtering stages is eligible to be selected block winner by the algorithm described in \cite{dftws}. There will only be one winner selected from this list, which means there is no guarantee to be selected block winner, even when a miner has correctly solved and uploaded the block problem solution.\\\\
In order for the second filtering stage (determining the most common solution) to be effective, it should not be possible for a malicious actor to create many node identities to influence what will be determined to be the most common solution. In order to solve this potential issue, a node will have to prove its real-world identity to the RA before being allowed to join the permissioned blockchain network.\\
The following outlines the steps it takes to participate in the network as a miner:
\begin{enumerate}\label{nodeIdentity}
	\item Generate Ed25519 private key
	\item Derive public key and libp2p node ID
	\item Prove real-world identity to RA and register with public key and libp2p node ID (RA will not know any miner's private key)
	\item RA will now broadcast a signed message that announces that a new node with the specified public key has joined the blockchain
\end{enumerate}
This will not only make it harder to create fake identities on the blockchain and try to perform a Sybil attack, but will also allow for a punishment mechanism where miners that repeatedly upload invalid solutions can be banned from the blockchain. This further improves the effectiveness of the most common solution approach, because it makes the assumption that there are more good nodes than malicious nodes more realistic.\\\\
All in all, combining the sub-problem matching with the most common solution approach allows the RA to probabilistically filter out false solutions without having to solve the entire block problem.
The solution verification is based on the notion ``\verb|fidete, sed verificate|'' which means ``\verb|trust, but verify|'': Trust refers to the assumption that at least some miners will act in good faith and upload correct solutions and verify refers to the sub-problem matching that allows verifying the correctness of certain sub-solutions.
\subsection{Block reward}
The distinguishing characteristic of the architecture described in this publication, compared to existing volunteer computing approaches such as e.g. BOINC projects, is that tradable tokens can be used to provide an incentive to participate in the blockchain. Similar to Bitcoin protocol specifications \cite{bitcoin}, tokens can only be created by solving block problems.\\
In our proposed blockchain, the amount of rewarded tokens will be halved every $1460$ blocks in order to create artificial token scarcity. This not only prevents token inflation but also is built on the notion that a currency without scarcity can not have perceived value. The RA is expected to target a block time of 24 hours, which means that one new block is created (and one block winner is rewarded) every day. This allows them to send difficult block problems to the network while also guaranteeing a reasonable transaction throughput.\\
The initial block reward is $64$ tokens for block $1$ and it gets halved every $1460$ blocks. The token reward of block $n \in \mathbb{N}$ can now be calculated using:\\\\
$R_n = 64 * \frac{1}{2}^{  \left \lceil{\frac{n}{1460}}\right \rceil  -1}$
\\\\This means the total token supply is limited to:\\\\
$\sum\limits_{n=1}^{\infty} 64 * \frac{1}{2}^{  \left \lceil{\frac{n}{1460}}\right \rceil  -1} = 186880$ tokens.\\\\
An advantage of Proof-of-Work variants over Proof-of-Stake or other approaches is that the former can be used to start a blockchain from scratch without allocating tokens in the genesis block.

\section{Challenges}\label{sec:challenges}
This section provides more information about challenges that had to be overcome to transition from traditional hash-based PoW to MC simulation-based PoUW for HEP.
\subsection{Identity control}
The verification approach described in section \hyperref[sec:verification]{"Verification of results"} partly relies on the  assumption that there exist honest miners that upload correct solutions, which will lead to collecting multiple solutions that have the same hash. The reason for the existence of this second filtering stage is that there is a chance that malicious miners could try to guess which sub-solutions will be used by the RA for the first filtering stage and pass it by luck, even though some of their sub-solutions are not correct. Such miners would usually not be able to pass the second filtering stage (because no other miner would have the same solution hash as them which means their solution hash will not be regarded to be the most common solution hash), but if malicious miners would be able to control many node identities, then they could submit the same false data with many different identities to have their false solution be viewed as the most common solution. In such a scenario, non-malicious nodes would not be eligible to be included on the winner selection list because their solution data would not pass the second filtering stage. In order to prevent such a Sybil attack, node identities must be controlled by the RA so make it infeasible to control multiple identities. The steps required to join the blockchain that are described in section \hyperref[sec:verification]{Verification of results} had to be introduced to prevent these kinds of Sybil attacks. The RA now is able to confirm real-world identities of miners which allows the implementation of a punishment protocol, in which bad actors can be banned from participating in the blockchain. As a result, nodes are less likely to cheat which has a positive effect on the most common solution approach.

\subsection{Data transmission bottlenecks}
Originally, the authors explored the idea of having the block problem be based on the analysis of real-world experimental data. However, this would force every miner to download large datasets before they could even start to perform useful work. Depending on the block time and the time available until a block problem expires, a miner with low bandwidth internet connection might not be able to start working on the actual problem before the block problem has expired. To prevent these issues, the block problem was adjusted to be Monte Carlo simulation-based. The advantage of this approach is that a difficult block problem can now be described with just a few parameters which prevents the initial data transmission bottlenecks. However, even with this approach the amount of data that results from executing the simulation task can be substantial (depends on MC simulation parameters like e.g. amount of events). This problem could be solved by combining the MC sim problem with an analysis task to create a few histograms or plots of interest to the RA. This way, the resulting data that needs to be uploaded to the RA can be minimized while the block problem itself can be difficult. By taking these steps data transmission bottlenecks can be circumvented. It also has the advantage that additional data for the MCPlots project can be generated this way.

\subsection{Spam attacks}
It is difficult to estimate the impact of spam attacks on the network before the implementation of gophy has been tested under real-world conditions. While it is true that the client software can be programmed to ignore requests from nodes whose identity was not registered to the RA, it also must be considered that determining whether a node is an accepted member of the blockchain or not takes a non-zero amount of computational power. It is difficult to estimate the impact of malicious nodes that flood e.g. the chaindb topic with data requests. It should be noted that registered nodes (as in their real-world identity has been proven to the RA) are disincentivized from participating in spam attacks as they can be banned from participating in the blockchain network by the RA in extreme cases. The feasibility and impact of various kinds of spam attacks remains to be evaluated.

\subsection{Token value}
The general idea of rewarding tradable tokens to miners is to provide an additional incentive (other than helping a real-world HEP experiment) to participate in the blockchain. But there are more approaches that could be taken to provide incentives to participate: One idea to make the collection of tokens worthwhile is to make use of gamification techniques (e.g. create a website that shows a ladder of the largest token holders to create competition between participants). This approach has successfully been employed by traditional volunteer computing projects. While in theory it would also be possible for the RA to provide a service that exchanges tokens for real-world currency at a specified rate (stablecoin), this approach is not recommended until the blockchain has proven itself to be resistant to attacks like e.g. spam attacks in practice.

\subsection{Computational effort of RA}
On the one hand, the RA must have local storage available to store uploaded data, data must be compared to determine most common solutions, certain sub-problems have to locally be calculated and servers that can handle certain amounts of network traffic (depends on network size and node activity) need to be run which can be costly. On the other hand, the RA can save money by instead of being forced to continuously invest in new hardware to satisfy their MC simulation requirements they can instead use the computational power of the blockchain network. As soon as the implementation is completed, the RA will be stress-tested under realistic real-world conditions. Only then estimations on possible cost savings of the RA can be made.

\section{Conclusion}
The authors have provided a brief overview of a novel PoUW blockchain architecture that has the goal of replacing hash-based PoW with simulation-based useful work. It has been shown that MC simulation tasks for HEP are a suitable block problem choice and the rationale behind architectural design decisions has been given. The role of the RA has been explained, and it has been made clear that a certain amount of centralization is required to preserve the usefulness of results. An overview of challenges that were overcome in the process was provided. The implementation of the software named gophy is in its final phase and is expected to be released and open-sourced within 2024. More detailed information about this project and an extensive documentation will follow.

\newpage
\bibliographystyle{alpha}
\bibliography{references}

\end{document}